\documentclass[fleqn,10pt]{wlscirep}
\usepackage[utf8]{inputenc}
\usepackage[T1]{fontenc}

\usepackage{here}

\usepackage{subcaption}

\usepackage{color}

\usepackage{soul}

\usepackage{booktabs,threeparttable}

\usepackage{bigstrut}
\usepackage{color}
\usepackage[normalem]{ulem}

\title{Recurrence in the evolution of air transport networks}

\author[1]{Kashin Sugishita}
\author[1, 2, 3, *]{Naoki Masuda}
\affil[1]{State University of New York at Buffalo, Department of Mathematics, Buffalo, 14260-2900, USA}
\affil[2]{State University of New York at Buffalo, Computational and Data-Enabled Science and Engineering Program, Buffalo, 14260-2900, USA}
\affil[3]{Waseda University, Faculty of Science and Engineering, 169-8555 Tokyo, Japan}
\affil[*]{naokimas@buffalo.edu}

\begin{abstract}
Changes in air transport networks over time may be induced by competition among carriers, changes in regulations on airline industry, and socioeconomic events such as terrorist attacks and epidemic outbreaks. Such network changes may reflect corporate strategies of each carrier. In the present study, we propose a framework for analyzing evolution patterns in temporal networks in discrete time from the viewpoint of recurrence. Recurrence implies that the network structure returns to one relatively close to that in the past. We applied the proposed methods to four major carriers in the US from 1987 to 2019. We found that the carriers were different in terms of the autocorrelation, strength of periodicity, and changes in these quantities across decades. We also found that the network structure of the individual carriers abruptly changes from time to time. Such a network change reflects changes in their operation at their hub airports rather than famous socioeconomic events that look closely related to airline industry. The proposed methods are expected to be useful for revealing, for example,  evolution of airline alliances and responses to natural disasters or infectious diseases, as well as characterizing evolution of social, biological, and other networks over time.

\end{abstract}
\begin{document}

\flushbottom
\maketitle
%
%
\thispagestyle{empty}


\section*{Introduction}

In the US, the Airline Deregulation Act was established in 1978, which removed the governmental controls over fares, routes, and market entries of new air carriers (carriers for short)\cite{brown1987politics}. After the establishment of the law, the formerly regional carriers and completely new carriers grew in the market and increased the level of competition among carriers\cite{goetz2009good}. The deregulation also caused the emergence of low cost carriers (LCCs). In contrast to full service carriers (FSCs), an LCC offers low fares without many traditional passenger services. The growth of LCCs, especially Southwest Airlines in the case of the US, which is the largest LCC in the world in terms of passenger traffic\cite{ren2020fare}, elicited a dramatic downward pressure on fares\cite{brueckner2013airline}. After the deregulation in the US, similar airline deregulations and subsequent elevated competition among carriers have been duplicated around the world including in Europe\cite{alderighi2012competition} and Asia\cite{zhang2008low}.

In addition to fierce competition among carriers, the airline industry has been exposed to a highly competitive environment partly owing to various socioeconomic events. For example, the 9/11 attacks in 2001 had significant impacts on the US airline industry. This event resulted in a negative demand shock of more than 30\%\cite{ito2005assessing}. The US airline industry lost nearly \$35 billion between 2001 and 2005 in total\cite{goetz2009good}. In fact, a number of carriers filed liquidation bankruptcies (so-called Chapter 7 bankruptcy) or reorganization bankruptcies (so-called Chapter 11 bankruptcy) including major carriers such as American Airlines, United Airlines, and Delta Air Lines\cite{ciliberto2012bankruptcy}. Moreover, the US airline industry has faced massive restructuring with a number of mergers such as the merger of American Airlines and US Airways, and of Delta Air Lines and Northwest Airlines\cite{steven2016mergers}. In Europe, the volcanic eruptions in Iceland in 2010 had a devastating effect on the European air transport network, preventing air travel throughout most of Europe for six days\cite{oroian2010eyjafjallajokull}. 
Overall, business environments in the airline industry have been tough on the global scale. For example, the International Air Transport Association (IATA) announced that year 2019 was the smallest in freight demand in the world since 2009\cite{IATA2020airline}. As of 2020, the airline industry is facing a financial crisis because of the COVID-19 pandemic. The Europe's biggest regional carrier of the time, Flybe, went bankrupt in March 2020, and many carriers may follow\cite{Harper2020airline}. 

The airline industry is highly susceptible to such socioeconomic and natural events, and these events may influence the structure of air transport networks. Regardless of the reason, air transport networks can be regarded as temporal networks in which nodes are airports and edges are direct flight connections\cite{rocha2017dynamics}; in various types of networks such as social and economic networks, nodes and/or edges change over time, constituting temporal networks \cite{holme2012temporal,holme2015modern,masuda2020guidance,holme2019temporal}.
In fact, air transport networks consist of multiple carriers. Each carrier has its own corporate strategies, which may change over time depending on evolution of the market and knowledge that the carrier gains by learning from databases (e.g., about customers)\cite{bieger2011airline}. These time-varying corporate strategies of carriers may affect the evolution of their network structure. 

A number of studies have analyzed structures of air transport networks. Most of these studies have examined static properties of air transport networks such as the degree distribution, average path length, clustering coefficient, assortativity, and centrality\cite{li2003structural, guimera2004modeling, guimera2005worldwide, guida2007topology, bagler2008analysis, xu2008exploring, cheung2012complex}. In recent years, several studies have investigated evolution of air transport networks. 
For example, in an air transport network in Brazil, the number of nodes and edges decreased from 1995 to 2006, whereas the number of passengers and the amount of cargo increased during the same period\cite{da2009structural}.
In contrast, the connectivity of an air transport network in China increased from 1930 to 2012\cite{wang2014evolution}.
These and other\cite{rocha2017dynamics,dai2018evolving,sun2015temporal,azzam2013accelerated} studies are concerned with the evolution of air transport networks at the country, continental, or global level and do not distinguish networks by carriers. In fact, each carrier may be evolving in different manners, reflecting its own corporate strategies. Because carriers compete for a limited traffic demand, some carriers expanding their networks may cause other carriers to shrink their networks. Even if the structure of an entire network combining different carriers is relatively stable\cite{lin2014evolving}, each carrier may be drastically changing its network structure over time. To date, time-dependent structure of air transport networks at the carrier level is not well understood\cite{lordan2014study}, while a few studies have investigated static properties of airline networks by carriers\cite{han2009network, reggiani2009network}.

In the present study, we propose a novel framework for analyzing evolution of air transport networks. In particular, we examine temporal air transport networks from the viewpoint of recurrence. Recurrence implies that the structure of the network returns to one close to that of the same network in the past. Recurrent plots and recurrence quantification analysis are a set of data analysis methods to quantify recurrence, which was originally proposed for time series \cite{marwan2007recurrence} and recently extended to the case of temporal networks \cite{masuda2019detecting, lopes2020recurrence}. While we do not employ recurrence plots or recurrence quantification analysis in the exact sense in the present study, we will use the idea of recurrence to examine when air transport networks approach their past structure relatively closely as compared to in other times. We apply the proposed framework to four major carriers in the US, i.e., American Airlines (AA), United Airlines (UA), Delta Air Lines (DL), and Southwest Airlines (WN), from 1987 to 2019. In particular, we ask how the network structure of different carriers responds to major events such as mergers, which airports are involved in major changes in the network structure, and how different carriers may be different in the extent of recurrence and periodicity of the network structure.

\section*{Methods}

\subsection*{Data set and construction of monthly air transport networks}
We use a data set available at the US Bureau of Transportation Statistics. The data are based on the reports made by certified carriers that account for at least one percent of the domestic scheduled passenger revenue. By definition, a certified carrier holds an operating certificate issued by the US Secretary of Transportation. The data set contains the origin and destination airports of each flight, departure and arrival times, operating carriers, and so on. We downloaded monthly data between October 1987 and August 2019\cite{BTS2020air}. 

We focus on air transport networks for the four most major carriers in the US in terms of revenue, i.e., AA, UA, DL, and WN. According to the US Bureau of Transportation Statistics, domestic revenue passenger-miles of each of these four carriers in 2019 is over 100 billions. Alaska Airlines, ranked fifth, earned domestic revenue passenger-miles less than 50 billions in 2019. For each carrier, we construct monthly snapshot networks from October 1987 to August 2019, in which a node is an airport. We connect two nodes by an edge if there is at least one direct commercial flight between the two airports in the month. For simplicity, we assume that each snapshot network is unweighted and undirected.

\subsection*{Distance matrix}
To measure the dissimilarity between two snapshot networks, we define the normalized network distance (network distance for short) between two networks $G$ and $G^{\prime}$ as
\begin{equation}
d(G,G^{\prime})= 1-\frac{M(G\cap G^{\prime})}{\sqrt{M(G)M(G^{\prime})}}\,,
\label{eq:normalized_network_distance}
\end{equation}
where $M(G)$ and $M(G^{\prime})$ are the numbers of edges in $G$ and $G^{\prime}$ respectively, and $M(G\cap G^{\prime})$ is the number of edges that $G$ and $G^{\prime}$ have in common. Network distance $d$ ranges between $0$ and $1$. The distance matrix for a carrier is a $t_{\max} \times t_{\max}$ symmetric matrix of which the $(t, t')$th entry is given by $d(G_t, G_{t'})$, 
where $G_t$ is the network at month $t$ and $t_{\max}$ is the number of months observed. If $d(G_t, G_{t'})$ is small, where $t < t'$, we regard that the network at month $t'$ approximately recurs to that at month $t$. As we will show, the distance matrices based on Eq.~\eqref{eq:normalized_network_distance} show clear horizontal and vertical lines that correspond to abrupt changes in the network structure, which are synchronous with the sudden increases in the number of nodes and edges upon major mergers. We decided to use Eq.~\eqref{eq:normalized_network_distance} because we did not observe such abrupt changes in the distance matrices calculated from three other network distance measures (see Supplementary Fig.~S1 online).

\subsection*{Statistical analysis}
To measure differences in the distance matrices, we conduct statistical tests on distributions of $d$. First, we use the Kruskal-Wallis test, which is an nonparametric statistical test to assess the differences among three or more groups\cite{kruskal1952use}. We used ``Kruskal'' function in R (ver. 3.6.3). Second, as a post-hoc test, we use the Bonferroni-corrected Mann-Whitney U test \cite{mann1947test} to compare all pairs of groups. We used ``wilcox.test'' function in R (ver. 3.6.3). In addition to the $p$ values, we measure its effect size $r$, which is the $Z$ value obtained from the ``wilcox.test'' function divided by the square root of the number of observations\cite{rosenthal1994parametric}. According to a standard, the effect size is said to be strong, medium, or small when $r>0.5$, $r>0.3$, or $r>0.1$, respectively\cite{cohen2013statistical}.

\subsection*{Autocorrelation function and power spectral density}
To measure the similarity between air transport networks across time, we define an autocorrelation function (ACF) for a given carrier by
\begin{equation}
\mbox{ACF}(\tau)=1-\frac{1}{t_{\max}-\tau}\sum_{t=1}^{t_{\max}-\tau} d(G_{t},G_{t+\tau})\,,
\label{eq:ACF}
\end{equation}
where $\tau$ is the time lag. Note that Eq. \eqref{eq:ACF} is not a proper autocorrelation function. However, in analogy to the Wiener–Khinchin theorem\cite{wiener2019cybernetics}, we define a power spectral density (PSD) by the Fourier transform of the ACF as follows: 
\begin{equation}
\mbox{PSD}(f)=\int_{-\infty}^{\infty} \mbox{ACF}(\tau)e^{-2\pi if\tau}{\rm d}\tau\,,
\label{eq:PSD}
\end{equation}
where $f$ is the frequency. The PSD shows the power of periodic variations. 

\subsection*{State detection in temporal air transport networks}
We identify states of the temporal air transport networks for each carrier based on its distance matrix. Consider a sequence of $t_{\max}$ static networks. To assign a state to each of the $t_{\max}$ snapshot networks, we apply a hierarchical clustering algorithm to the $t_{\max} \times t_{\max}$ distance matrix\cite{masuda2019detecting}. We used ``linkage'' and ``fcluster'' functions in ``scipy'' module in Python. To define the distance between clusters, we used ``ward'' in the ``linkage'' function. The hierarchical clustering provides partitions of the monthly networks into $C$ discrete states, where the number of states, $C$, ranges between $1$ and $t_{\max}$. We adopt the value of $C\ (2 \le C \le t_{\max})$ that maximizes the Dunn's index, $D$ \cite{dunn1974fuzzy}, defined by
\begin{equation}
D=\frac{\min_{1 \leq c \neq c' \leq C} \min_{G_i \in c{\rm th\ state}, G_j \in c'{\rm th\ state}} d(G_i,G_j)}{\max_{1 \leq c'' \leq C} \max_{G_{i'}, G_{j'} \in c''{\rm th\ state}} d(G_{i'},G_{j'})}\,.
\label{eq:dunn}
\end{equation}
The numerator on the right-hand side of Eq.~\eqref{eq:dunn} represents the smallest distance between two states among all the pairs of states. The denominator represents the largest diameter of the state among all the states.

\section*{Results}
\subsection*{Evolution of the numbers of nodes and edges}
We show the evolution of the numbers of nodes and edges for each carrier in Fig.~\ref{fig:nodes_edges}. The three FSCs (i.e., AA, UA, and DL) experienced the Chapter 11 bankruptcies. This is a form of bankruptcy that permits reorganization under the bankruptcy laws. A debtor proposes a plan of reorganization, and it can exit from the bankruptcy if it completes the proposed plan. The dashed lines in Fig.~\ref{fig:nodes_edges} represent the times of the Chapter 11 bankruptcies (shown in red) and exits from them (shown in yellow). Additionally, we show the times of the 9/11 attacks and mergers by the dotted-dashed lines and dotted lines, respectively.

The three FSCs show some common patterns. Between the middle of the 1990s and around 2010, the numbers of nodes and edges largely decreased. After the 9/11 attacks, the three carriers experienced the bankruptcies. Then, after they exited from the bankruptcy, they merged with other carriers. Upon the mergers, the number of edges almost doubled suddenly. There was no notable change in the numbers of nodes and edges in response to the bankruptcies and the exits from them. In contrast, WN, which is an LCC, has never experienced bankruptcy. The numbers of nodes and edges in the WN network have roughly monotonically increased over the two decades despite the 9/11 attacks and the merger. In the next sections, we further study the structural evolution of the four carriers.

\subsection*{Distance matrix for each carrier}

\subsubsection*{Visualization and quantification of the distance matrices}
\begin{figure}[H]
\centering
\includegraphics[width=0.88\linewidth]{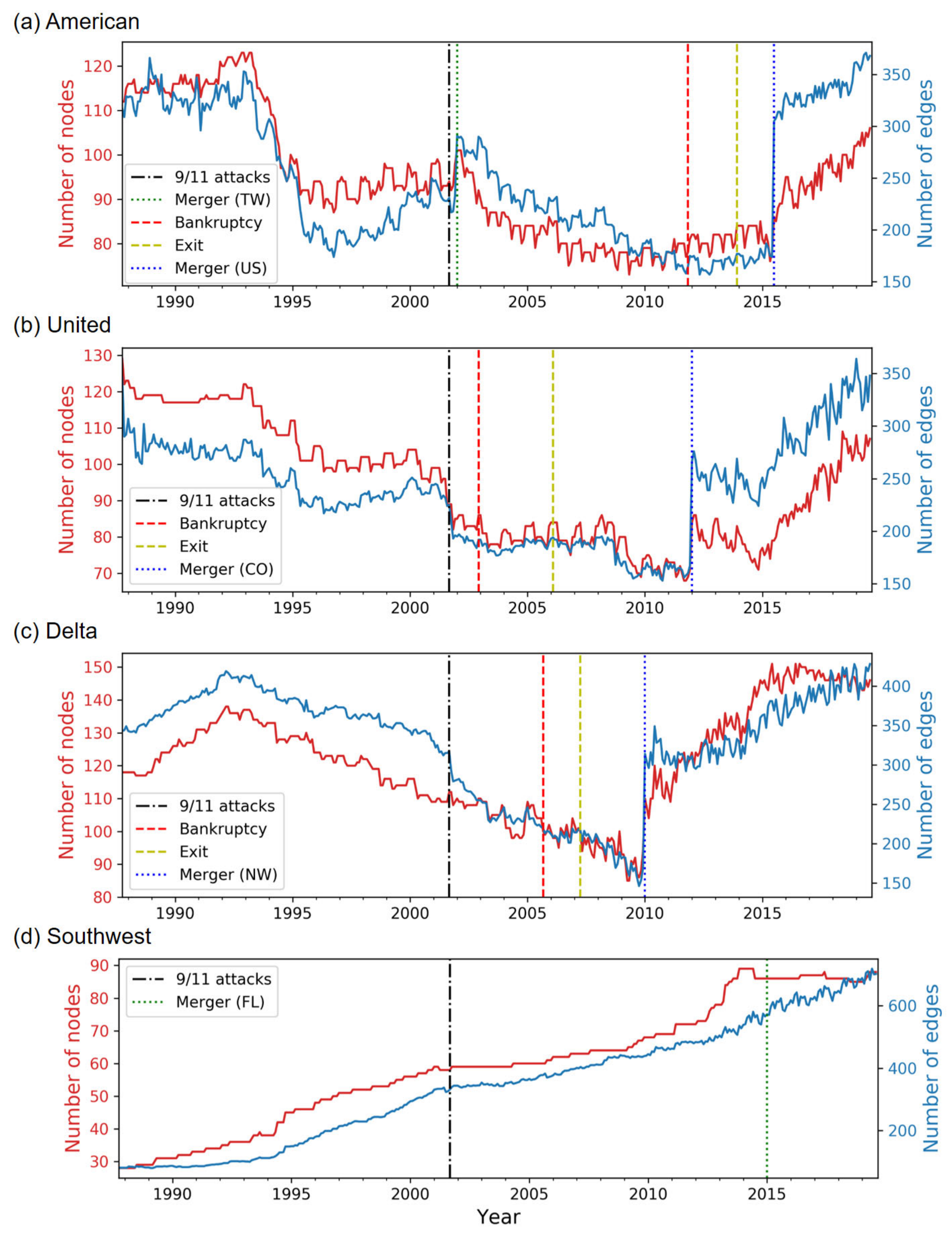}
\caption{Evolution of the numbers of nodes and edges over time. (a) American Airlines, (b) United Airlines, (c) Delta Air Lines, and (d) Southwest Airlines.}
\label{fig:nodes_edges}
\end{figure}

\begin{figure}[htb]
\centering
\includegraphics[width=0.99\linewidth]{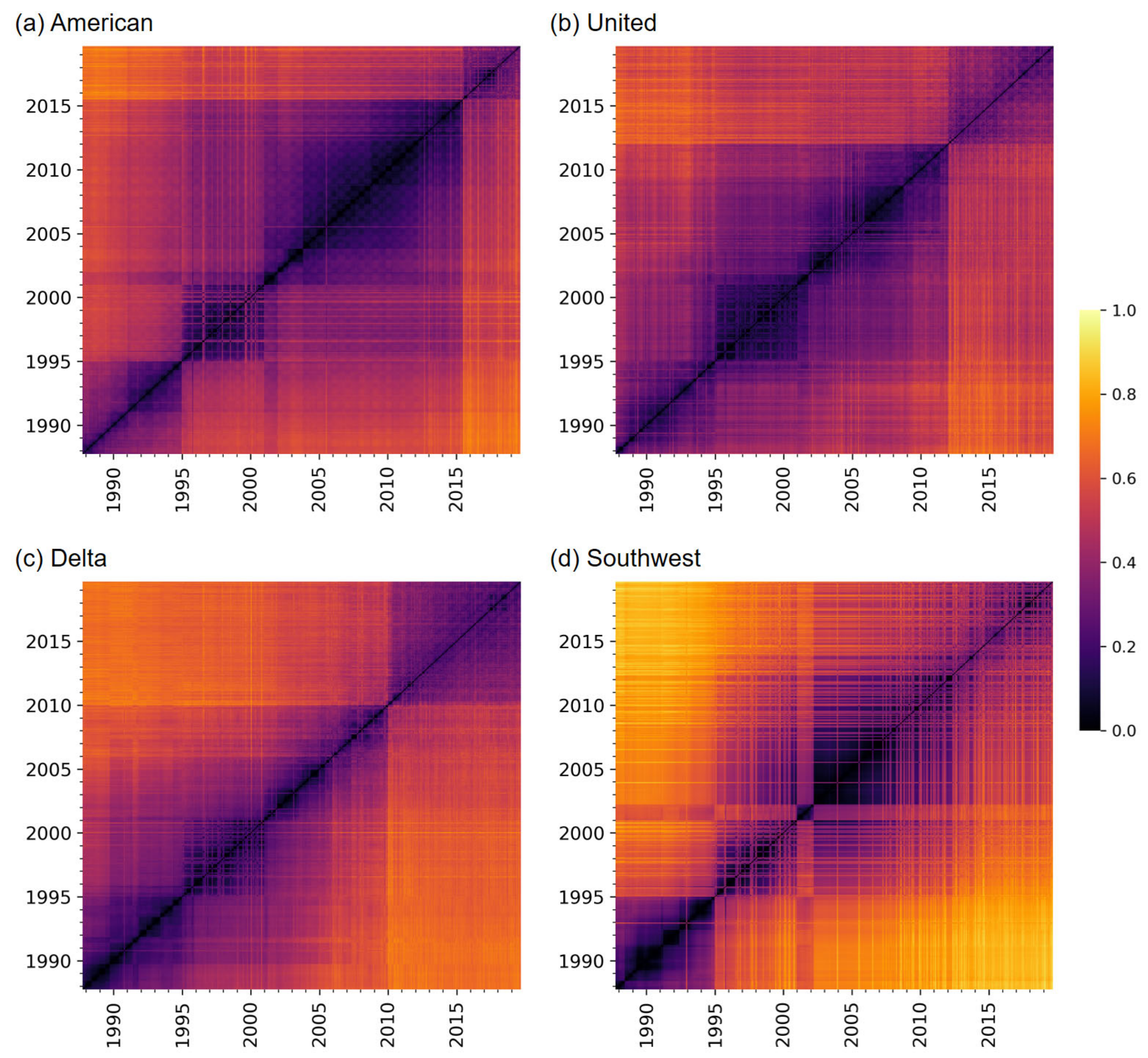}
    \caption[]
    {Network-distance matrices. (a) American Airlines, (b) United Airlines, (c) Delta Air Lines, and (d) Southwest Airlines.} 
    \label{fig:distance_matrices}
\end{figure}

In Fig.~\ref{fig:distance_matrices}, we show the distance matrices for the four carriers. First, we notice that there are clear horizontal and vertical lines in the distance matrices. These lines indicate abrupt changes in the network structure. For example, the figure suggests that the three FSCs considerably changed their network structure upon the mergers after they exited from the bankruptcies (i.e., the merger between AA and US Airways (US) in 2015 in Fig.~\ref{fig:distance_matrices}a, that between UA and Continental Airlines (CO) in 2012 in Fig.~\ref{fig:distance_matrices}b, and that between DL and Northwest Airlines (NW) in 2010 in Fig.~\ref{fig:distance_matrices}c). These results are consistent with the approximately two-fold sudden increase in the numbers of edges shown in Figs.~\ref{fig:nodes_edges}a-c.  However, the distance matrices additionally indicate that the network structure after the merger is considerably different from that before the merger for each FSC. For example, the DL network after the merger with NW in 2010 is considerably different from that in the late 1990s, although the numbers of nodes and edges are similar at the two time points (Fig.~\ref{fig:nodes_edges}c).

Second, there are checkerboard patterns in the distance matrices, for example, around year 2010 for AA (Fig.~\ref{fig:distance_matrices}a). The checkerboard patterns imply seasonal periodicity in the network structure. These patterns are also consistent with the annual periodic fluctuations in the number of nodes and edges visible in Fig.~\ref{fig:nodes_edges}a. Note that the classical recurrence plots\cite{marwan2007recurrence}, in which the distance values are binarized into 0 or 1, would not uncover patterns of the distance matrix either at relatively large distance values (e.g., the sudden changes in response to the post-bankruptcy mergers) or at relatively small distance values (e.g., the checkerboard patterns). In other words, recurrence plots with a small threshold value for binarization will not have the long horizontal and vertical lines marking major mergers because relatively large distance values that contribute to generation of these lines are all binarized to 1. By contrast, recurrence plots with a large threshold value will miss the checkerboard patterns because relatively small distance values that contribute to the generation of the checkerboard patterns are all binarized to 0. In our framework, we analyze the original values of the network distances without binarizing them, which enables us to reveal various patterns in the distance matrices without varying a threshold parameter.

Third, we notice that the distance matrix for WN (Fig.~\ref{fig:distance_matrices}d) has many bright horizontal and vertical stripes after 1995. They suggest that WN has frequently changed their network, but for a short duration each time such as a single month. 

\begin{table}[thb]
\centering
\caption{\label{tab:Statistical_analysis_1} Results of the Bonferroni-corrected Mann-Whitney U test to compare each pair of carriers in terms of $d$. ***: $p<0.001$.}
\begin{threeparttable}
\begin{tabular}{|c|c|}
\hline
   & Effect size, $r$        \\ \hline
AA vs UA & $0.0129^{***}$ \\ \hline
AA vs DL & $0.211^{***}$  \\ \hline
AA vs WN & $0.325^{***}$  \\ \hline
UA vs DL & $0.225^{***}$ \\ \hline
UA vs WN & $0.337^{***}$    \\ \hline
DL vs WN & $0.162^{***}$      \\ \hline
\end{tabular}
\footnotesize
  \end{threeparttable}
\end{table}

\begin{figure}[H]
\centering
\includegraphics[width=0.39\linewidth]{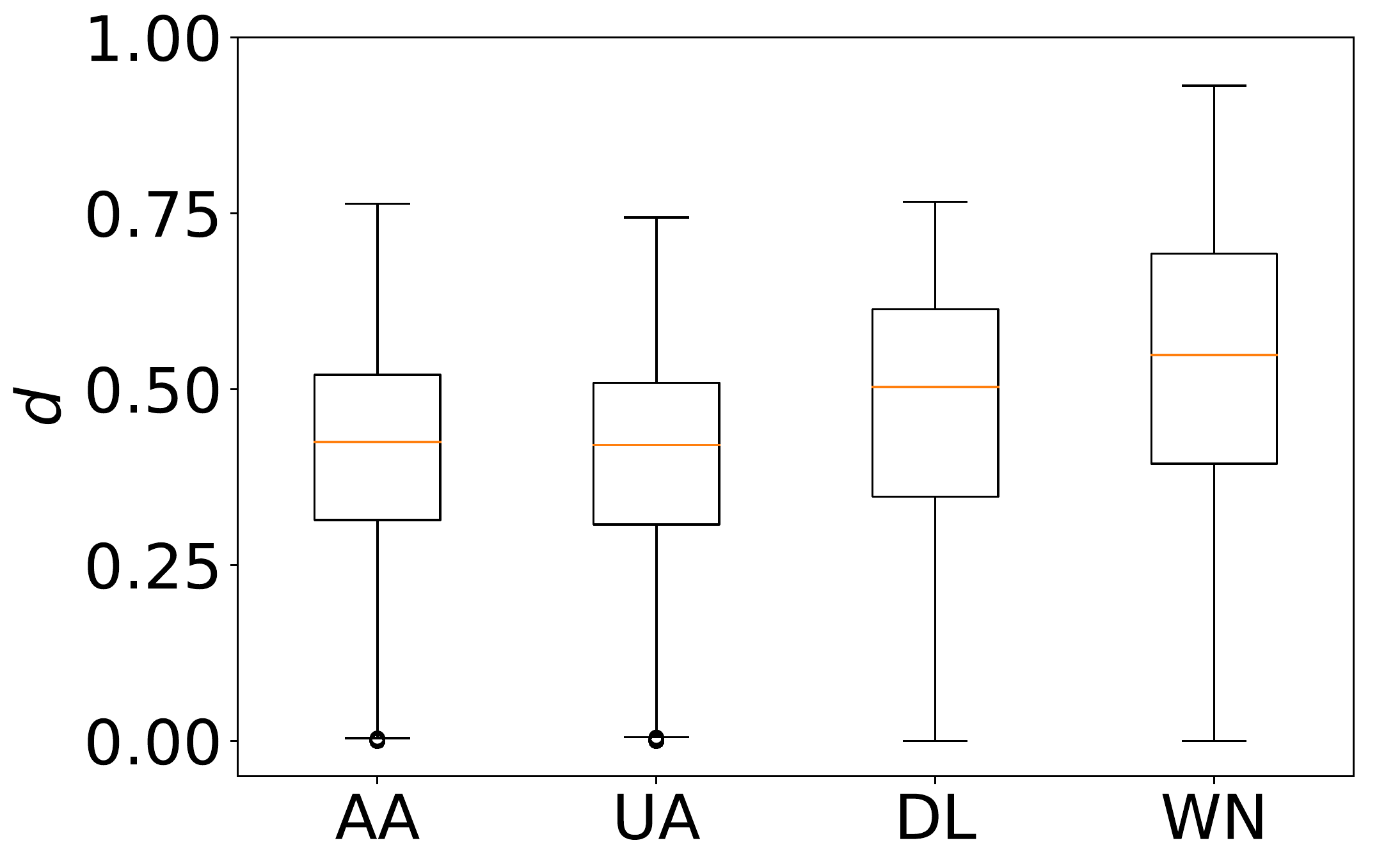}
\caption{The distribution of the network distance $d$. The box plots show five-number summaries of the distribution: first quartile ($Q_1$), median, third quartile ($Q_3$), minimum without outliers ($Q_1-1.5 \, \times \, IQR$), and maximum without outliers ($Q_3+1.5 \, \times \, IQR$), where IQR $=Q_3-Q_1$. Open circles represent outliers.}
\label{fig:distance_all_period}
\end{figure}

In Fig.~\ref{fig:distance_all_period}, we show the box plots of the distribution of the network distance between pairs of months. Each box plot summarizes each distance matrix shown in Fig.~\ref{fig:distance_matrices}. To be quantitative, we carry out statistical analyses on these distributions. First, we carry out the Kruskal-Wallis test to find that there is a significant difference among the four distributions shown in Fig.~\ref{fig:distance_all_period} ($p<0.001$). Second, as a post-hoc test, we run the Bonferroni-corrected Mann-Whitney U test to compare all six pairs of carriers. Table~\ref{tab:Statistical_analysis_1} indicates that each pair of carriers is significantly different, partly owing to large sample sizes. According to a standard practice, two pairs, AA-WN and UA-WN, are different with medium effect sizes ($r>0.3$), and three pairs, DL-WN, AA-DL, and UA-DL, are different with small effect sizes ($r>0.1$). The effect size for pair AA-UA is only 0.0129. These results support our casual observation with Fig.~\ref{fig:distance_matrices}, i.e., WN shows unique patterns and has kept changing the network structure more considerably than the other three carriers. Additionally, DL changed its structure significantly more than AA and UA did over time.

\subsubsection*{Periodicity}
\begin{figure}[h]
\centering
\includegraphics[width=0.99\linewidth]{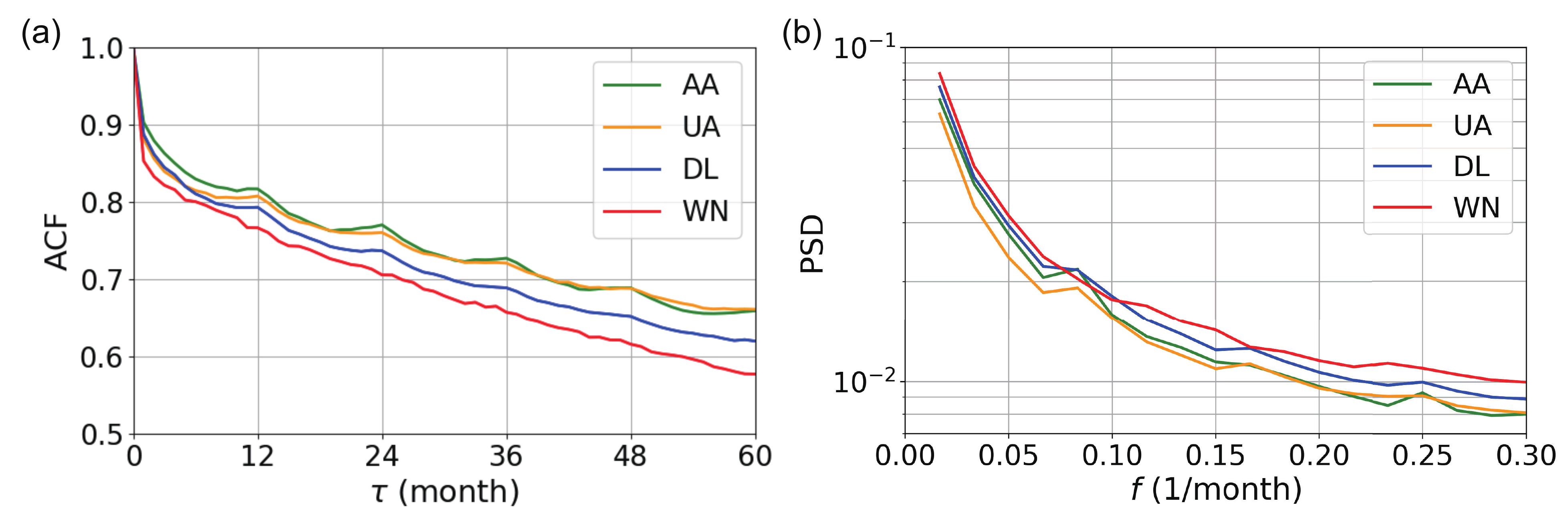}
\caption{Analysis of autocorrelation and periodicity. (a) ACFs and (b) PSDs for the four carriers.}
\label{fig:ACF_PSD_four_air_carriers}
\end{figure}

In Fig.~\ref{fig:ACF_PSD_four_air_carriers}a, we show the ACFs for the four carriers, calculated from the distance matrices shown in Fig.~\ref{fig:distance_matrices}. This figure indicates that networks closer in time are more similar to each other. We also observe that the ACF for WN is the smallest for all the lag $\tau >0$, confirming that the WN network tends to vary over time to a larger extent than the other carriers. In addition, Fig.~\ref{fig:ACF_PSD_four_air_carriers}a reconfirms that the DL network tends to vary over time more than the AA and UA networks do.

\begin{figure}[h]
\centering
\includegraphics[width=0.38\linewidth]{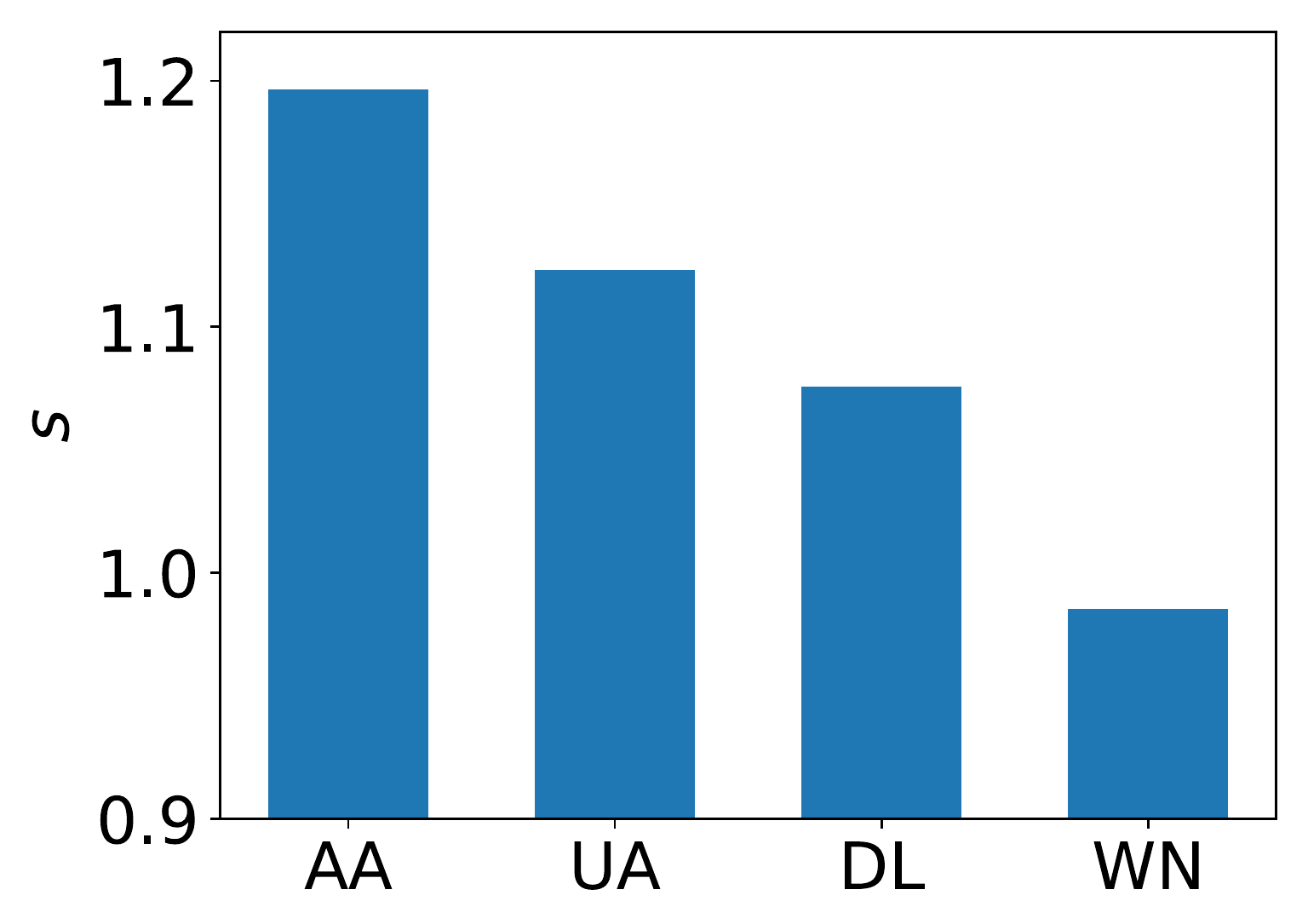}
\caption{Strength of annual periodicity in the air transport networks.}
\label{fig:strength_periodicity_four_airlines}
\end{figure}

The ACFs for the FSCs have local maxima at multiples of 12 months, suggesting annual periodic patterns. To quantify the strength of annual cycles, we show the PSDs for the four carriers in Fig.~\ref{fig:ACF_PSD_four_air_carriers}a. The PSDs have local maxima, corresponding to an annual cycle, for AA and UA. Here we define the strength of the annual periodicity, denoted by $s$, at $f=1/12$ [1/month] by comparing $\mbox{PSD}(1/12)$ with the middle point of the two neighbors, $\mbox{PSD}(1/15)$ and $\mbox{PSD}(1/10)$, as follows:
\begin{equation}
s=\frac{\mbox{PSD}(1/12)}{(\mbox{PSD}(1/15)+\mbox{PSD}(1/10))/2}\,.
\label{eq:strength}
\end{equation}
A value of $s$ larger than 1 suggests some periodicity with frequency $f=1/12$. We show $s$ for each carrier in Fig.~\ref{fig:strength_periodicity_four_airlines}. The figure suggests that AA has the strongest annual periodicity, followed by UA and DL. The value of $s$ for WN is less than 1, suggesting the lack of periodicity.  

\subsubsection*{State transitions in temporal air transport networks}
In this section, we examine states of temporal air transport networks for each carrier.
In Figs.~\ref{fig:nodes_edges} and \ref{fig:distance_matrices}, we observed that the network structure for the three FSCs drastically changed upon the mergers after they exited from the bankruptcy. If we apply our algorithm to detect states of temporal networks to the entire observation, we detect two states for each FSC, one state corresponding to the time window before the merger and another state after the merger. However, a visual inspection of Fig.~\ref{fig:distance_matrices} suggests that we may benefit from distinguishing multiple states for the time period before the merger. Therefore, for each FSC, here we apply the hierarchical clustering algorithm to detect state dynamics to the distance matrix of the temporal network restricted to before the post-bankruptcy merger. We regard the snapshot networks after the post-bankruptcy merger as one distinct state. For WN, we applied the algorithm to detect system states to the entire observation window because the WN network has not considerably changed upon the merger with AirTran Airways (FL) in 2015 up to our visual inspections of Figs.~\ref{fig:nodes_edges}d and \ref{fig:distance_matrices}d.

\begin{figure}[bth]
\centering
\includegraphics[width=0.83\linewidth]{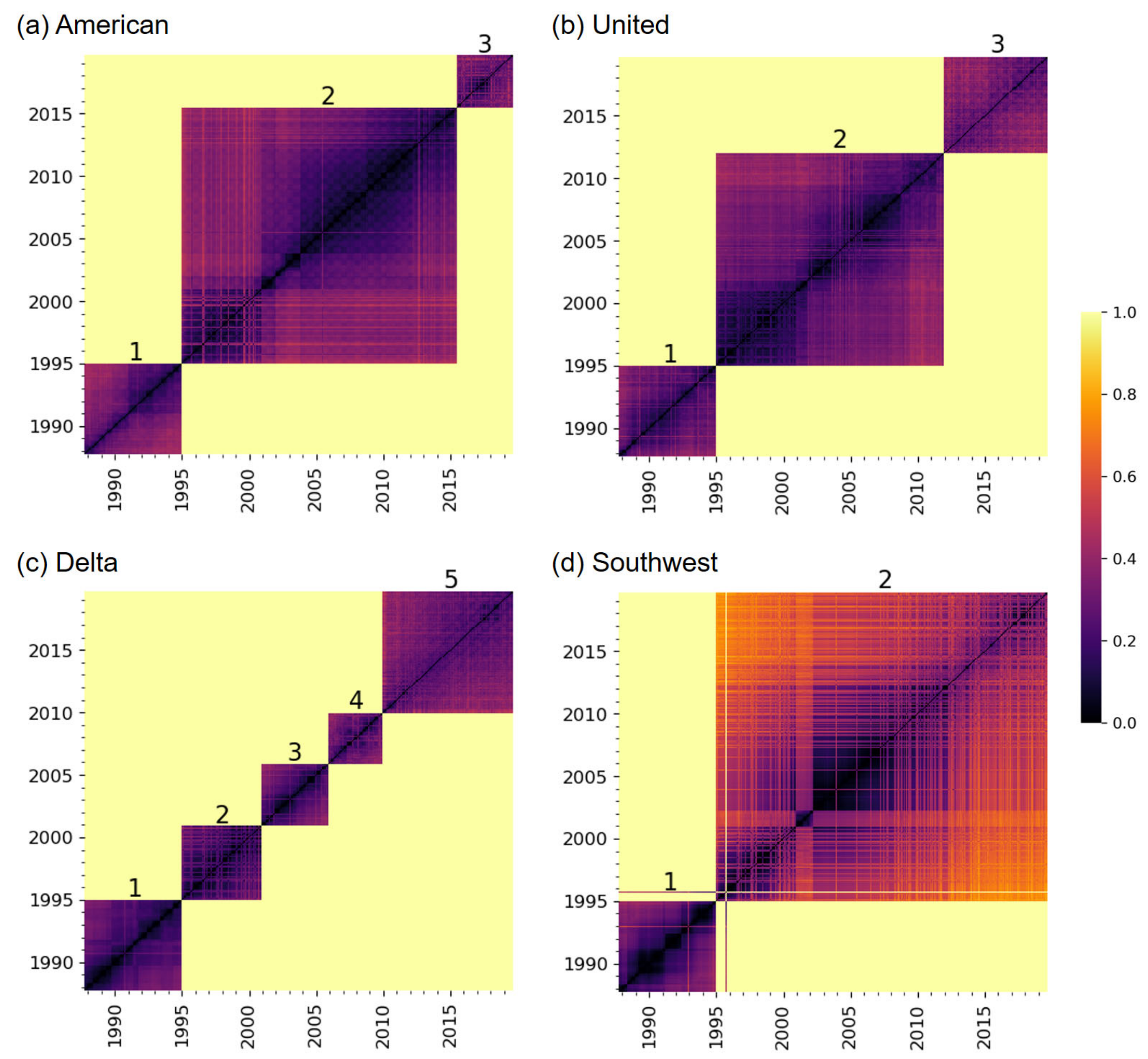}
\caption{System-state dynamics of temporal air transport networks. (a) American Airlines, (b) United Airlines, (c) Delta Air Lines, and (d) Southwest Airlines. To aid visual understanding, we show the network distance on a color scale if and only if the two snapshot networks belong to the same state. Otherwise, the corresponding cell is shown in yellow.}
\label{fig:detected_states}
\end{figure}

We show the detected state dynamics for the four carriers in Fig.~\ref{fig:detected_states}. For all the four carriers, we detected a change point at January 1995. We identified three states for AA and UA, and five states for DL. The detected changes for DL from state 2 to 3 and state 3 to 4 are located at January 2001 and December 2005, respectively. For WN, we detected two states. Differently from the other carriers, state 1 for the WN network is not consecutive in time, i.e., state 1 reappears in October 1995 for a month after the state transition on January 1995. We notice that the detected change point at December 2005 for DL is three months after the bankruptcy. However, except for this instance and the state after the post-bankruptcy mergers for the FSCs, the detected change points coincide with neither the 9/11 attacks (i.e., September 2001), the bankruptcies (i.e., AA: November 2011, UA: December 2002), nor the exits from the bankruptcies (i.e., AA: December 2013, UA: February 2006, DL: April 2007).

\begin{figure}[H]
\centering
\includegraphics[width=\linewidth]{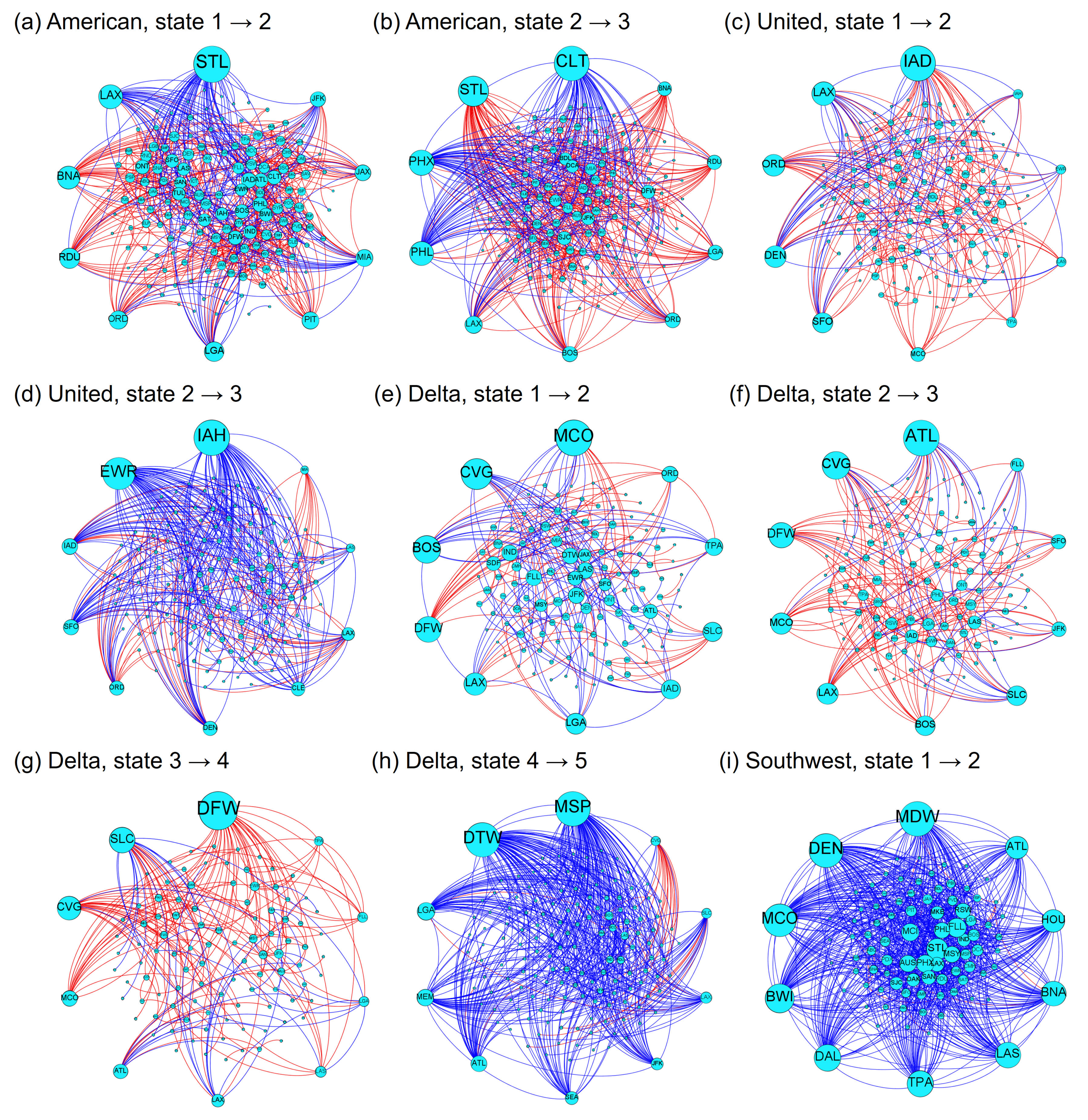}
\caption{Deleted edges (shown in red) and added edges (shown in blue) for each pair of the consecutive states. In each panel, we placed 10 nodes with the largest number of added and deleted edges in total in the peripheral part of the visualization.}
\label{fig:change_networks}
\end{figure}

To understand differences in the network structure across the states, we visualize deleted and added edges for each pair of consecutive states. We define deleted edges as the edges that exist at least once in the older state and never exist in the newer state. The added edges are defined as edges that never exist in the older state and exist at least once in the newer state. We show the results in Fig.~\ref{fig:change_networks}. The deleted and added edges are shown in red and blue, respectively. 

Figure~\ref{fig:change_networks}a indicates that, in transition to state 2, AA withdrew edges especially at BNA (Nashville International Airport) and RDU (Raleigh-Durham International Airport) and added edges at STL (St. Louis Lambert International Airport) and LAX (Los Angeles International Airport). These results are consistent with the fact that AA closed two hub airports, BNA and RDU, in the middle of the 1990s and merged with Trans World Airlines (TW) in 2002 whose major airports were STL and LAX. Figure~\ref{fig:change_networks}b indicates that, in transition to state 3, AA withdrew edges by the largest number at STL and added many edges at CLT (Charlotte Douglas International Airport), PHX (Phoenix Sky Harbor International Airport), and PHL (Philadelphia International Airport). These results are consistent with the fact that AA ceased to use STL as a hub around 2010 and merged with US whose major airports were CLT, PHX, and PHL.

Figure~\ref{fig:change_networks}c indicates that, in transition to state 2, UA changed many routes connecting to IAD (Washington Dulles International Airport), which has been a hub airport for UA. Figure~\ref{fig:change_networks}d indicates that, in transition to state 3, UA added edges by the largest numbers at IAH (George Bush Intercontinental Airport) and EWR (Newark Liberty International Airport). This event originated from the merger with CO whose major airports had been IAH and EWR. 

Figures~\ref{fig:change_networks}e and \ref{fig:change_networks}f indicate that, in transition from state 1 to 2 and from state 2 to 3, DL changed many edges connecting to some airports such as MCO (Orlando International Airport), CVG (Cincinnati/Northern Kentucky International Airport), and ATL (Hartsfield–Jackson Atlanta International Airport). Figure~\ref{fig:change_networks}g shows that, in transition to state 4, DL withdrew edges by the largest number at DFW (Dallas/Fort Worth International Airport). Figure~\ref{fig:change_networks}h is consistent with the merger with NW whose major airports were MSP (Minneapolis–Saint Paul International Airport) and DTW (Detroit Metropolitan Airport). 

Figure~\ref{fig:change_networks}i indicates that, in transition to state 2, WN expanded the network by adding many edges at some airports such as MDW (Midway International Airport), DEN (Denver International Airport), and MCO.

\subsubsection*{State-dependent periodicity}
In the previous section, we documented differences in the network structure across the states through visualization. We expect that the different states are systematically different in some quantitative aspects. Therefore, in this section, we quantitatively examine the temporal variability of the network and possible annual periodicity of network changes, which themselves may change on the timescale of decades. We measure the ACF and PSD for each state and each carrier. Note that a lower ACF suggests a larger temporal variability of time series\cite{murray2014hierarchy,cavanagh2016autocorrelation,watanabe2019atypical}. Because state 1 for WN is not consecutive in time, we exclude the single isolate month in state 1 (i.e., October 1995) and nine isolated months in state 2 (i.e., from January to September 1995) in the following analysis in the following analysis.

\begin{figure}[H]
\centering
\includegraphics[width=0.97\linewidth]{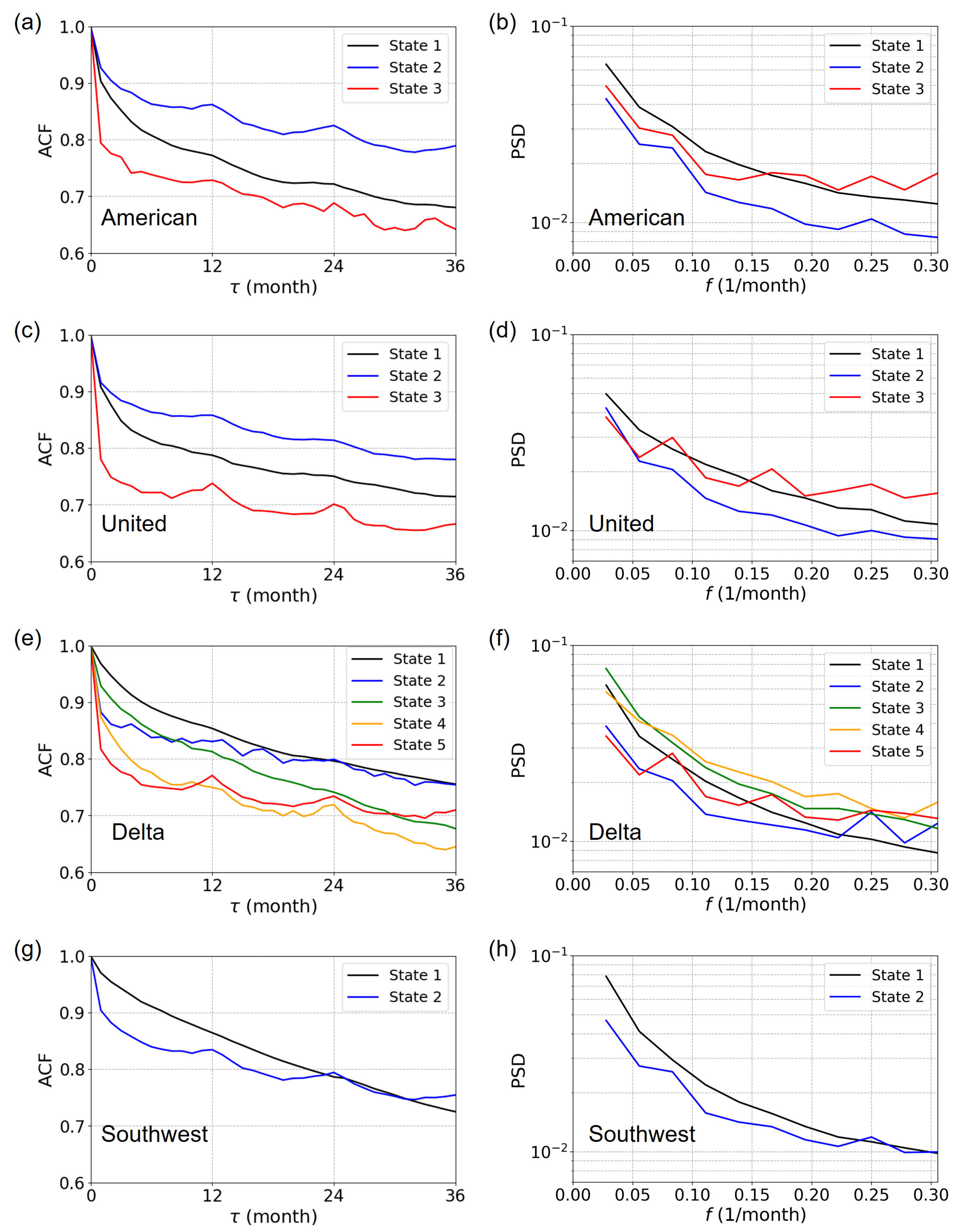}
\caption{ACFs and PSDs for each network state. (a) American Airlines, ACF. (b) American Airlines, PSD. (c) United Airlines, ACF. (d) United Airlines, PSD. (e) Delta Air Lines, ACF. (f) Delta Air Lines, PSD. (g) Southwest Airlines, ACF. (h) Southwest Airlines, PSD.}
\label{fig:ACFs_PSDs_each_state}
\end{figure}

We show the ACFs and PSDs for each state and each carrier in Fig.~\ref{fig:ACFs_PSDs_each_state}. For AA and UA, the ACF is the largest in state 2, followed by state 1, and then by state 3, for all the lag values (Fig.~\ref{fig:ACFs_PSDs_each_state}a,c). This result indicates that the temporal variability is smaller (i.e., the ACF is larger) in state 2, which roughly spans from 1995 to 2013 for both carriers. In contrast, the ACF for DL and WN tends to be smaller at later times, at least up to $\tau=2$ years of the lag (Fig.~\ref{fig:ACFs_PSDs_each_state}e,g). This result indicates that the temporal variability becomes larger over time for DL and WN.

\begin{figure}[h]
\centering
\includegraphics[width=0.65\linewidth]{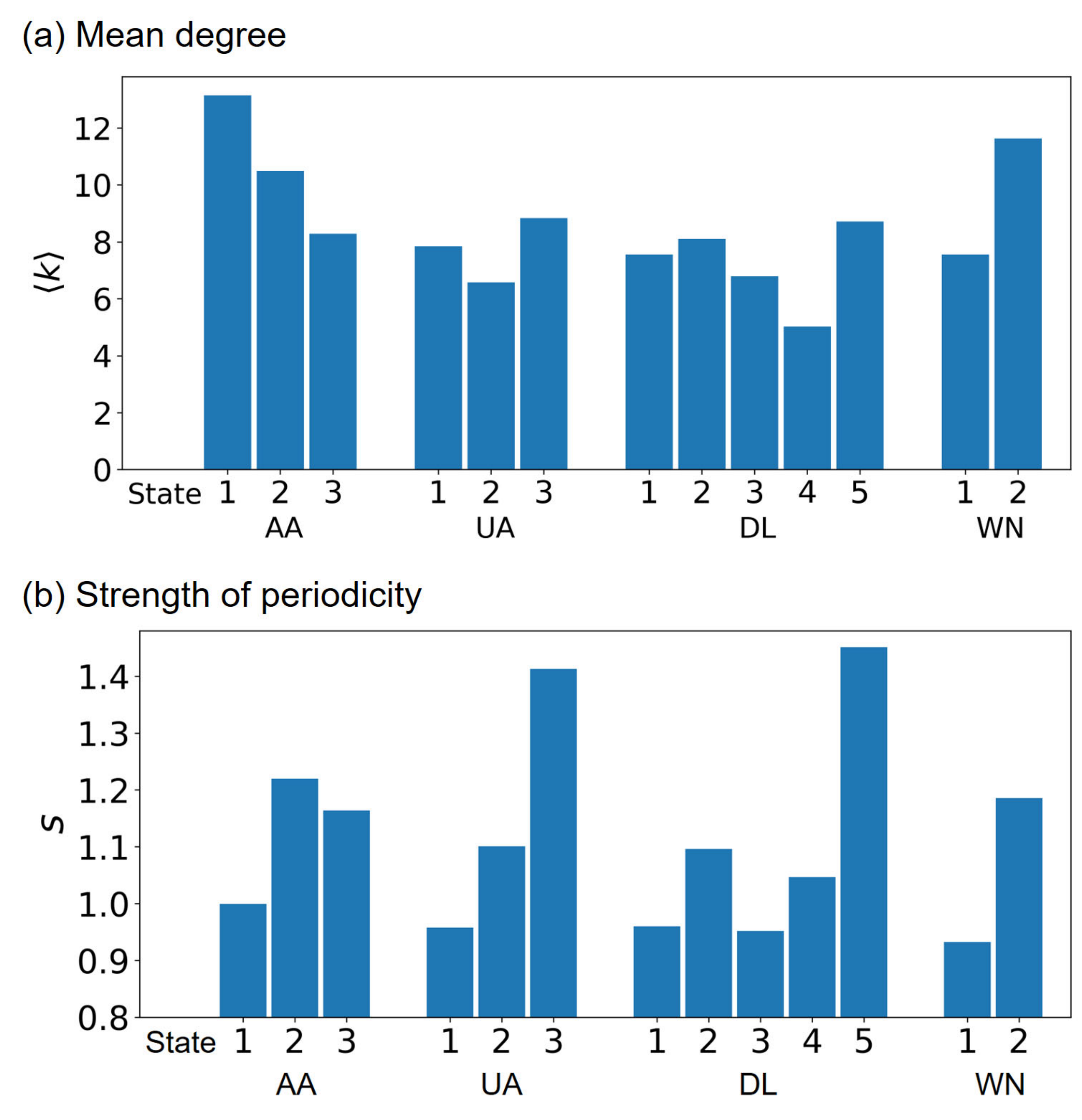}
\caption{Mean degree and strength of periodicity in each state.}
\label{fig:strength_periodicity_all_states}
\end{figure}

The PSDs for some combinations of the state and carrier have a local maximum at $f=1/12$. We measure the strength of periodicity in the same manner as Eq. \eqref{eq:strength}, i.e., by comparing $\mbox{PSD}(1/12)$ with the middle point of the two neighbors, i.e., $\mbox{PSD}(1/18)$ and $\mbox{PSD}(1/9)$. It should be noted that the largest lag assumed in Fig.~\ref{fig:ACFs_PSDs_each_state} is 30 as opposed to 50 in Fig.~\ref{fig:ACF_PSD_four_air_carriers}, which results in different $f$ values for the two neighbors of $f=1/12$ (i.e., $f=1/18$ and $1/9$) as compared to those for Fig.~\ref{fig:ACF_PSD_four_air_carriers}. We show the mean degree, $\langle k \rangle$, and the strength of periodicity, $s$, for the different states and carriers in Fig.~\ref{fig:strength_periodicity_all_states}. The figure suggests that both $\langle k\rangle$ and $s$ depend on the states but in different manners. For example, the periodicity is much larger in the final state in the UA and DL networks, which the state dependence of $\langle k\rangle$ does not explain. The figure further indicates that all the four carriers did not have a clear periodicity before 1995 (corresponding to state 1 for each carrier). After 1995, all carriers strengthened the periodicity. In the latest states, UA and DL have stronger periodicity than AA and WN.

\subsection*{Network distance between carriers}
\begin{figure}[H]
\centering
\includegraphics[width=\linewidth]{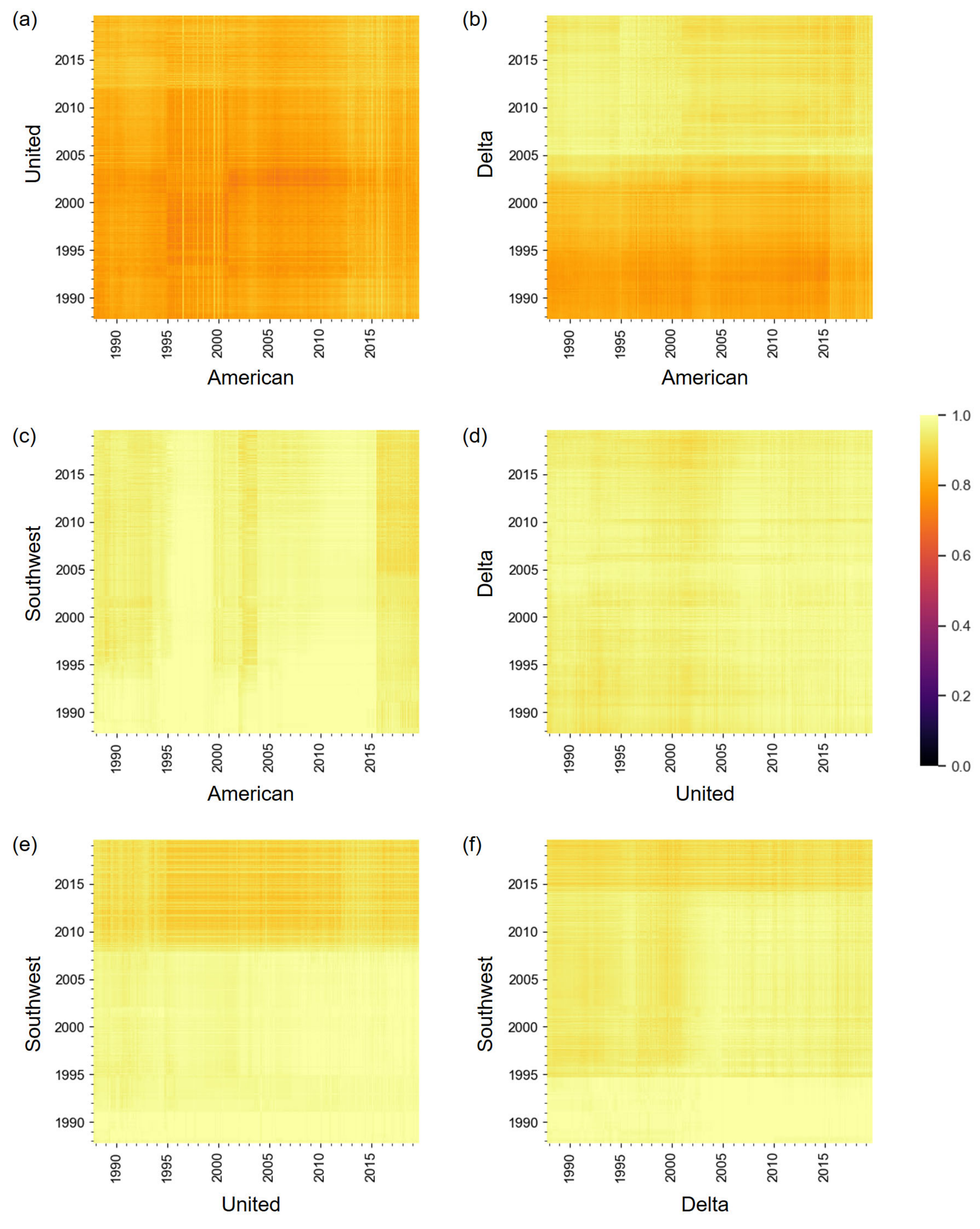}
\caption{Network distance between different carriers over time. (a) American Airlines - United Airlines, (b) American Airlines - Delta Air Lines, (c) American Airlines - Southwest Airlines, (d) United Airlines - Delta Air Lines, (e) United Airlines - Southwest Airlines, and (f) Delta Air Lines - Southwest Airlines.}
\label{fig:distance_matrices_different_airlines}
\end{figure}

In this section, we examine similarity in the network structure across different carriers. We calculate the distance matrix for each pair of carriers by measuring the normalized network distance between a network at month $t$ for a carrier and a network at month $t'$ for another carrier, where $1\le t,\  t' \le t_{\max}$. The distance matrices in Fig.~\ref{fig:distance_matrices_different_airlines} indicate that the four carriers are all dissimilar to each other; note that the distance values range between 0 and 1 by definition. Nevertheless, the AA and UA networks are apparently more similar to each other than the other carrier pairs are on average. Another observation is that there are some visible horizontal and vertical lines in the figure (e.g., a horizontal line in 2005 in Fig.~\ref{fig:distance_matrices_different_airlines}b, a vertical line in 2015 in Fig.~\ref{fig:distance_matrices_different_airlines}c, and a horizontal line in 2008 in Fig.~\ref{fig:distance_matrices_different_airlines}e).

\begin{table}[h]
\caption{\label{tab:Statistical_analysis_2} Results of the Bonferroni-corrected Mann-Whitney U test to compare all 15 pairs of carriers in terms of $d$. ***: $p<0.001$.}
  \begin{minipage}[t]{.33\textwidth}
    \begin{center}
     \begin{tabular}{|c|c|}
\hline
   & Effect size, $r$        \\ \hline
AA-UA vs AA-DL & $0.512^{***}$ \\ \hline
AA-UA vs AA-WN & $0.866^{***}$  \\ \hline
AA-UA vs UA-DL & $0.866^{***}$  \\ \hline
AA-UA vs UA-WN & $0.855^{***}$ \\ \hline
AA-UA vs DL-WN & $0.865^{***}$    \\ \hline
\end{tabular}
    \end{center}
  \end{minipage}
  \hfill
  \begin{minipage}[t]{.33\textwidth}
    \begin{center}
\begin{tabular}{|c|c|}
\hline
   & Effect size, $r$        \\ \hline
AA-DL vs AA-WN & $0.742^{***}$ \\ \hline
AA-DL vs UA-DL & $0.693^{***}$  \\ \hline
AA-DL vs UA-WN & $0.585^{***}$  \\ \hline
AA-DL vs DL-WN & $0.621^{***}$ \\ \hline
AA-WN vs UA-DL & $0.304^{***}$    \\ \hline
\end{tabular}
    \end{center}
  \end{minipage}
  \hfill
    \begin{minipage}[t]{.33\textwidth}
    \begin{center}
     \begin{tabular}{|c|c|}
\hline
   & Effect size, $r$        \\ \hline
AA-WN vs UA-WN & $0.134^{***}$ \\ \hline
AA-WN vs DL-WN & $0.243^{***}$  \\ \hline
UA-DL vs UA-WN & $0.151^{***}$  \\ \hline
UA-DL vs DL-WN & $0.0112^{***}$ \\ \hline
UA-WN vs DL-WN & $0.0551^{***}$    \\ \hline
\end{tabular}
    \end{center}
  \end{minipage}
  %
\end{table}

\begin{figure}[h]
\centering
\includegraphics[width=0.52\linewidth]{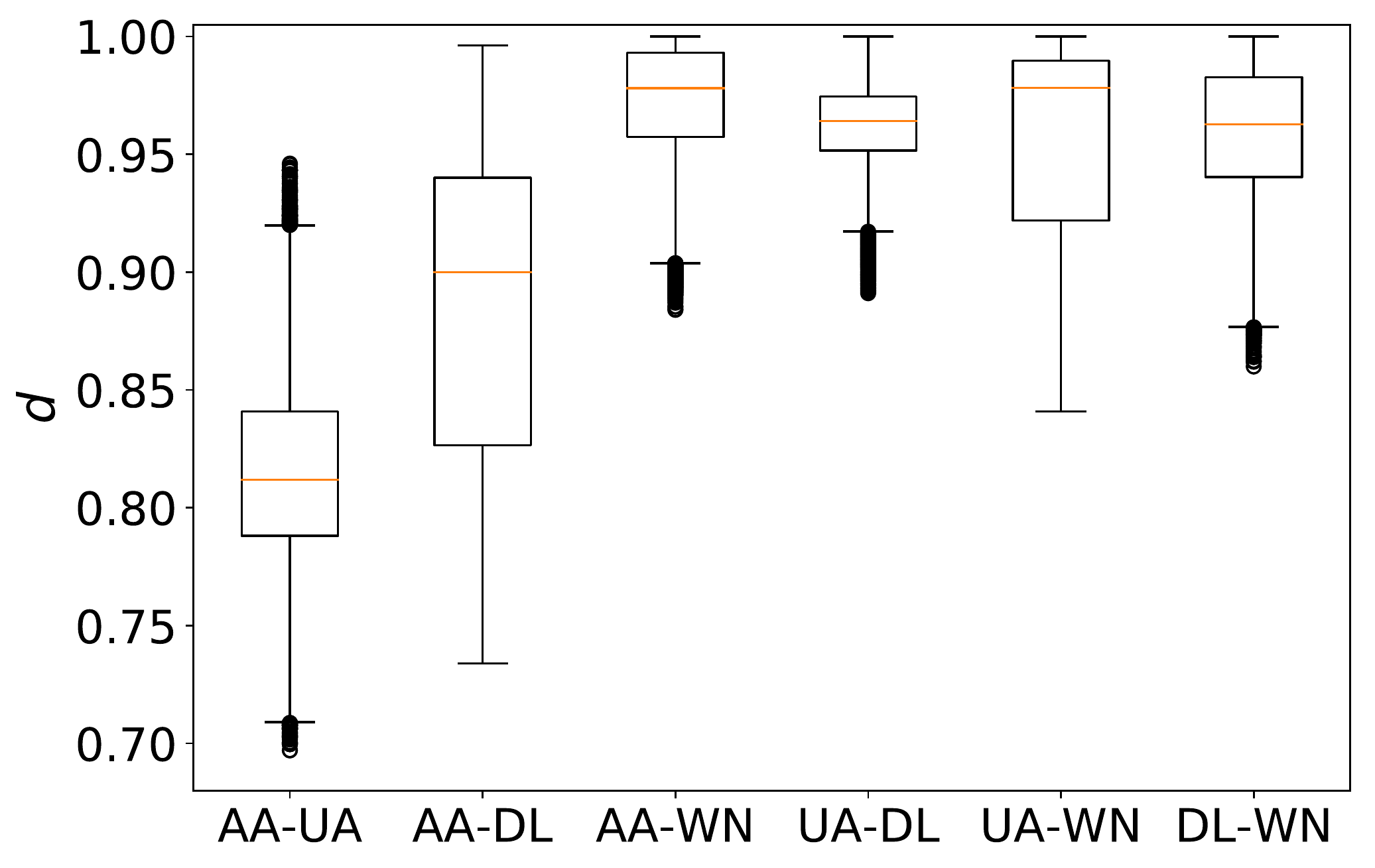}
\caption{Distribution of the network distance for the pairs of carriers.}
\label{fig:box_2}
\end{figure}

The box plots for the entries of the network distance matrix for each pair of carriers are shown in Fig.~\ref{fig:box_2}. Consistent with Fig.~\ref{fig:distance_matrices_different_airlines}a, the AA and UA networks seem to be more similar to each other on average than the other pairs. We then carried out the Kruskal-Wallis test and found that there was a significant difference in these distributions ($p<0.001$). The Bonferroni-corrected Mann-Whitney U test  as a post-hoc test revealed that there were significant differences in the $p$-values for all 15 pairs with various effect sizes (see Table~\ref{tab:Statistical_analysis_2} for the statistical results). In particular, the network distance values for AA-UA are smaller than those for any other carrier pairs with strong effect sizes ($r>0.5$). The network distance values for AA-DL are also smaller than the other carrier pairs except for the AA-UA pair with strong effect sizes ($r>0.5$). These results indicate that these two carrier pairs are more similar than the other carrier pairs.

\begin{figure}[H]
\centering
\includegraphics[width=0.98\linewidth]{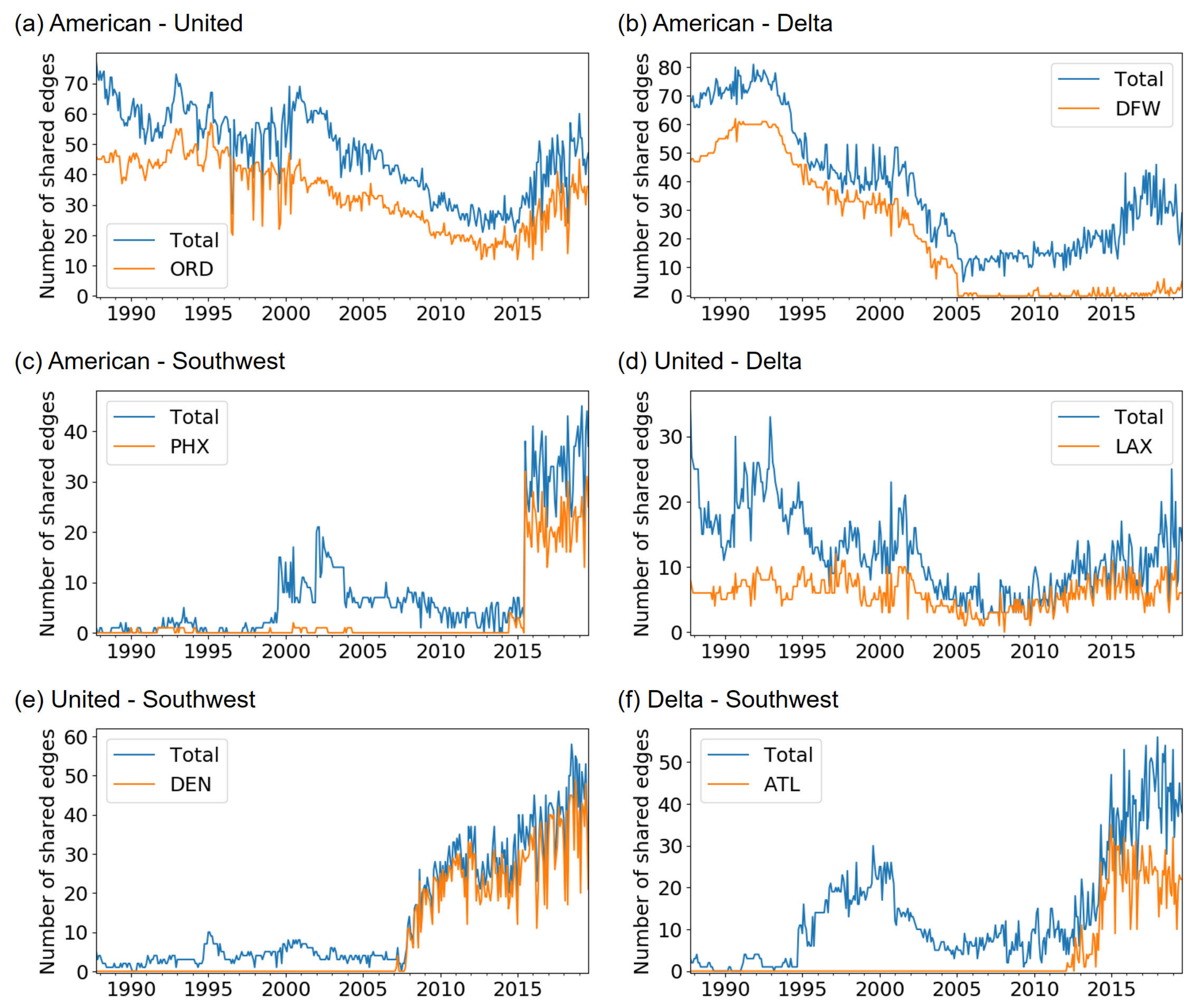}
\caption{ Evolution of the number of edges shared by pairs of carriers. We show the total number of shared edges and the number of shared edges connecting to major airports for each pair of carriers. (a) American Airlines versus United Airlines, (b) American Airlines versus Delta Air Lines, (c) American Airlines versus Southwest Airlines, (d) United Airlines versus Delta Air Lines, (e) United Airlines versus Southwest Airlines, and (f) Delta Air Lines versus Southwest Airlines. ORD: O'Hare International Airport, DFW: Dallas/Fort Worth International Airport, PHX: Phoenix Sky Harbor International Airport, LAX: Los Angeles International Airport, DEN: Denver International Airport, and ATL: Hartsfield–Jackson Atlanta International Airport.}
\label{fig:common_edges}
\end{figure}

To search for the origin of the observed similarities and dissimilarities between carrier pairs, we show the time series of the total number of edges shared by two carriers in each month and the number of such shared edges connecting to some major airports in Fig.~\ref{fig:common_edges}. Fig.~\ref{fig:common_edges}a indicates that the relatively large similarity in the entire observation period between AA and UA originates from the shared edges connecting to ORD (O'Hare International Airport), which has been a major hub for both AA and UA. Fig.~\ref{fig:common_edges}b indicates that AA and DL became more dissimilar around 2005 because DL closed a hub at DFW, which we also observed in Fig.~\ref{fig:change_networks}g. Figure~\ref{fig:common_edges}c indicates that AA and WN became more similar around 2015. This is because AA merged with US, one of whose major airports had been PHX, as we also observed in Fig.~\ref{fig:change_networks}b. Figure~\ref{fig:common_edges}d indicates that most of the edges that UA and DL share originate from the operations at LAX. Figure~\ref{fig:common_edges}e,f indicates that WN became more similar to UA and DL because WN increased the number of edges at DEN and ATL, which we also observed in Fig.~\ref{fig:change_networks}i. These results indicate that large part of the similarity in each carrier pair originates from operations at a single airport shared by the two carriers.

\section*{Discussion}
We proposed a framework for analyzing evolution patterns in temporal air transport networks. We applied the proposed methods to the four largest carriers in the US, i.e., AA, UA, DL, and WN, between 1987 and 2019. We found that their overall evolution patterns were different. Specifically, our analysis suggested that WN changed the network structure more than the other carriers and that DL changed the network structure more than AA and UA. Their evolution patterns are also different in terms of the temporal variability and the strength of periodicity. Furthermore, we found that each carrier experienced abrupt changes in their network structure. We also found that the temporal variability and the strength of periodicity tended to increase in the latest years.

All the four carriers drastically changed their network structure in January 1995. In addition, they strengthened the periodicity after 1995. To the best of our knowledge, this change point is not related to any major socioeconomic and industrial events. The reason for this change point is unclear. We also showed that the three FSCs (i.e., AA, UA, and DL) substantially changed their network structure upon the post-bankruptcy mergers. On the other hand, we did not detect notable changes in the network structure upon the 9/11 attacks, the bankruptcies, and the exits from the bankruptcies, except for a change point for DL three months after the bankruptcy. There are different possible mechanisms that may have produced these results. First, these events may have triggered drastic changes in the network structure, but on a shorter time scale than a month. Analysis of weekly or daily snapshot networks may be fruitful; our methods are not selective of the time scale, and detailed flight data for the US carriers with a time resolution of minutes is available at the US Bureau of Transportation Statistics\cite{BTS2020air}. Second, only the number of flights between airports may have primarily changed in response to such events. Although we have assumed unweighted networks in the present article, if one modifies the network distance measure to accommodate weighted networks, the methods are directly applicable to the case of weighted temporal networks. These research directions warrant future work.

We showed that WN, which is an LCC, has shown particularly distinct patterns of evolution over years as compared to the FSCs, such as a consistent growth trend in terms of the number of nodes and edges, relatively fast changes in the network structure, and lack of periodicity when the data from all years are combined. To generalize these results to differences between LCCs and FSCs, one needs more thorough analyses of different carriers, such as LCCs and FSCs in Europe and Asia, where LCCs have been rapidly growing. We opted to study the US case in the present study because of the availability of the data.

Our data-analytic framework is also applicable to different contexts. For example, an application to airline alliances (e.g., Oneworld, SkyTeam, and Star Alliance) may reveal influences of entries and exits of carriers on the networks of airline alliances over years. Another direction is to study influences of natural disasters on the structure of air transport networks. For example, ash cloud caused by the volcanic eruption in Iceland in 2010 affected the structure of the European air transport networks\cite{wilkinson2012vulnerability, dunn2016increasing}. Another possible application is to study influences of epidemic outbreaks including the COVID-19 pandemic, which is ongoing as of mid 2020. Obviously, carriers have been slashing down flights (corresponding to the edge weight) and edges due to the drop in the demand. In addition to this, some carriers may be strategically maintaining their network structure to be as similar as possible, or perhaps the converse, to that before COVID-19. Some carriers may also be changing the similarity of their network to other carriers' networks. Finally, the proposed methods including the analysis of periodicity are not specific to air transportation networks such that they are applicable to temporal networks in general.




\bibliography{ref}



\section*{Acknowledgements (not compulsory)}
K.S. acknnowledges the financial support by JSPS. N.M. acknnowledges the financial support by AFOSR European Office. 

\section*{Author contributions statement}
N.M. conceived and designed the research; K.S. collected and analyzed the data; K.S. and N.M. wrote the paper.

\section*{Additional information}

\end{document}